\documentclass[twocolumn,american,english,prb]{revtex4-1}
\usepackage[T1]{fontenc}
\usepackage[latin9]{inputenc}
\usepackage{amsmath}
\usepackage{amssymb}
\usepackage{graphicx}

\makeatletter

\pdfpageheight\paperheight
\pdfpagewidth\paperwidth

\@ifundefined{textcolor}{}
{%
 \definecolor{BLACK}{gray}{0}
 \definecolor{WHITE}{gray}{1}
 \definecolor{RED}{rgb}{1,0,0}
 \definecolor{GREEN}{rgb}{0,1,0}
 \definecolor{BLUE}{rgb}{0,0,1}
 \definecolor{CYAN}{cmyk}{1,0,0,0}
 \definecolor{MAGENTA}{cmyk}{0,1,0,0}
 \definecolor{YELLOW}{cmyk}{0,0,1,0}
}

\usepackage{babel}

\usepackage{babel}

\makeatother

\usepackage{babel}
\begin{document}

\title{Renormalization and Scaling in Quantum Walks }

\author{Stefan Boettcher$^{1}$, Stefan Falkner$^{1}$, and Renato Portugal$^{2}$}

\affiliation{$^{1}$Department of Physics, Emory University, Atlanta, GA 30322;
USA\\
 $^{2}$ Laborat\'orio Nacional de Computa\c{c}\~ao Cient{\'\i}fica, Petr\'opolis, RJ
25651-075; Brazil}
\begin{abstract}
We show how to extract the scaling behavior of quantum walks using the renormalization group (RG).
We introduce the method by efficiently reproducing well-known results on the one-dimensional
lattice. As a nontrivial model, we apply this method to the dual Sierpinski gasket and obtain its
exact, closed system of RG-recursions. Numerical iteration suggests that under rescaling the system
length, $L^{\prime}=2L$, characteristic times rescale as $t^{\prime}=2^{d_{w}}t$ with the exact walk
exponent $d_{w}=\log_{2}\sqrt{5}=1.1609\ldots$. Despite the lack of translational invariance, this
is very close to the ballistic spreading, $d_{w}=1$, found for regular lattices. However, we argue
that an extended interpretation of the traditional RG formalism will be needed to obtain scaling
exponents analytically. Direct simulations confirm our RG-prediction for $d_w$ and furthermore
reveal an immensely rich phenomenology for the spreading of the quantum walk on the gasket.
Invariably, quantum interference localizes the walk completely with a site-access probability that
declines with a powerlaw from the initial site, in contrast with a classical random walk, which
would pass all sites with certainty.
\end{abstract}
\maketitle

\section{Introduction}

Following Grover's work,\cite{Gro97a} it was shown that discrete-time
quantum walks\cite{AAKV01,SKW03,Bach2004562,AKR05,PortugalBook} can
access any chosen site of a regular lattice in two or more dimensions
at least in $O(\sqrt{N\ln N})$ steps. Such a quadratic speed-up over
classical $O(N)$ first-passage times is one of the promising aspects
of quantum computing, for instance, to search for items in an unsorted
list. The fundamental importance of search algorithms to databases
can not be overstated, especially for hierarchical networks without
translational invariance. We only need to mention Google's page-rank
algorithm, for which a quantum version was recently proposed.\cite{Paparo13}
As fundamental as the random walk is to the description of classical
diffusion and transport phenomena in physics\cite{Weiss94,Redner01}
or the mixing times of randomized algorithms in computer sciences,\cite{MM11}
the analogous quantum walk is rapidly rising in importance to describe
a range of phenomena. Already, there are a number of experimental
realizations of quantum walks, such as in waveguides,\cite{Perets08,Martin11}
photonics,\cite{Sansoni12,Crespi13} and optical lattices.\cite{Weitenberg11}
Therefore, classifying the \emph{physical} behavior of quantum walks,
their entanglement, localization, and interference effects in complex
environments, is interesting in its own right.

To date, only very few analytical means\cite{PortugalBook,Carteret03}
exist to describe the wealth of experimental and numerical observations.
Aside from path-integral methods, these are mostly based on using
a Fourier decomposition of the walk equation that presupposes a translational
or re-labeling symmetry between all sites. Accordingly, the general
quantum walk on a simple line is by now relatively well explored,\cite{AAKV01,SKW03,Bach2004562}
with a few forays into specific instances of two-dimensional lattices.\cite{MBSS02,AKR05,OPD06}
However, insistence on translational invariance leaves us with a limited
understanding of the full impact of quantum interference effects,
which are the origin of the quadratic speed-up in the spreading on
regular lattices. However, the range of studied systems remains too
narrow to assess -- let alone, predict -- how interference causes
any particular scaling. In addition, localization effects emerge as
soon as lattices possess loops\cite{IK05,Inui05,Falkner14a} or disorder.\cite{Shikano10}

Here, we develop the venerable real-space renomalization group
(RG)\cite{Wilson71,Goldenfeld,Pathria} to discover the long-range behavior of discrete-time quantum
walks in more complex geometries. We introduce RG for the simple line, where we show how to
reproduce the well-known ballistic spreading exponent, $d_{w}=1$, by extending the traditional fixed
point analysis into the complex plane.\cite{Boettcher13a} For quantum walks on the dual Sierpinski
gasket (DSG), iterating our exact recursions to $k=21$ generations, corresponding to a gasket with
$N\approx3^{21}\approx10^{10}$ nodes, shows that time $t$ rescales with base-line length $L$ as
$t\sim L^{d_{w}}$ with $d_{w}=1.16096\ldots=\log_{2}\sqrt{5}$, not quite ballistic but spreading
faster than a random walk on DSG,\cite{Weber10} for which $d_{w}^{RW}=\log_{2}5$. However, we find
that localization effects diminish the magnitude of the wave function almost everywhere by
$\left|\psi\right|\sim L^{-\beta}$ with $\beta=0.424(3)$, such that \emph{extensive} transport (that
can still reach the boundaries for increasing $L$) decays with a power of $L$ that is bounded
between $d_{w}$ and $d_{w}+2\beta$, and we have $1<d_{w}<d_{w}+2\beta<d_{w}^{RW}$. We test our
predictions with direct simulations on DSG with up to $k=12$ generations.

This paper is organized as follows. In the next section, we introduce a formulation of the walk
problem that allows to study classical and quantum walks on comparable footing. In
Sec.~\ref{sec:Renormalization-1dQW}, we apply the RG for walks, classical and quantum, on the simple
line. In Sec.~\ref{sec:Renormalization-DSG}, we study the RG for the dual Sierpinski gasket. In
Sec.~\ref{sec:RG-Fixed-Points-are}, we discuss the unusual aspects of the RG for quantum walks. In
Sec.~\ref{sec:Conclusions}, we conclude with a summary of our results and outline future work.

\section{Formulation of the Walk Problem\label{sec:Solution-of-QW-1}}

The generic master-equation for a discrete-time walk with a coin,
whether classical or quantum, is 
\begin{equation}
\left|\Psi\left(t+1\right)\right\rangle ={\cal U}\left|\Psi\left(t\right)\right\rangle ,\label{eq:MasterE}
\end{equation}
where the time-evolution operator (or propagator) is written as 
\begin{equation}
{\cal U}={\cal S}\left({\cal C}\otimes{\cal I}\right),\label{eq:Ushift}
\end{equation}
containing the ``shift'' operator ${\cal S}$ and the coin ${\cal C}$.
In the $d$-dimensional site-basis $\left|\vec{n}\right\rangle $,
we can describe the state of the system in terms of the site amplitudes
$\psi_{\vec{n},t}=\left\langle \vec{n}|\Psi\left(t\right)\right\rangle $,
simply the probability density to be at that site for a classical
walk, but representing in the quantum walk a vector in coin-space
with each component holding the amplitude for transitioning out of
site $\vec{n}$ along one of its links. Application of the coin ${\cal C}$
entangles these components, with subsequent redistribution of the
walk to neighboring sites by the shift operator ${\cal S}$, based
on those amplitudes. On the line, the shift operator for a homogeneous
nearest-neighbor walk is 
\begin{equation}
{\cal S}=\sum_{x}\left\{ P\otimes\left|x-1\right\rangle \left\langle x\right|+Q\otimes\left|x+1\right\rangle \left\langle x\right|+R\otimes\left|x\right\rangle \left\langle x\right|\right\} \label{eq:Shift}
\end{equation}
with the shift matrices $P$ and $Q$ for moving left or right, and
$R$ for staying in place, for instance, 
\begin{equation}
P=\left(\begin{array}{cc}
1 & 0\\
0 & 0
\end{array}\right),\quad Q=\left(\begin{array}{cc}
0 & 0\\
0 & 1
\end{array}\right),\quad R=\left(\begin{array}{cc}
0 & 0\\
0 & 0
\end{array}\right).\label{eq:PQRinit-1dQW}
\end{equation}
From Eq.~(\ref{eq:Ushift}), we get the propagator
\begin{equation}
{\cal U}  =  \sum_{x}\left\{ A\otimes\left|x-1\right\rangle \left\langle x\right|+B\otimes\left|x+1\right\rangle \left\langle x\right|+M\otimes\left|x\right\rangle \left\langle x\right|\right\} \label{eq:propagator-1}
\end{equation}
with $A=P{\cal C}$, $B=Q{\cal C}$, and $M=R{\cal C}$, and the unitary
coin-matrix ${\cal C}$, most generally,\cite{PortugalBook} 
\begin{equation}
{\cal C}=\left(\begin{array}{cc}
\sin\eta & e^{i\chi}\,\cos\eta\\
e^{i\vartheta}\,\cos\eta & -e^{i\left(\chi+\vartheta\right)}\,\sin\eta
\end{array}\right).\label{eq:QW1dCoin-1}
\end{equation}

In a quantum walk, the ``hopping'' operators $A$, $B$, and $M$
are constrained by the requirement of unitary propagation, ${\cal I}={\cal U}^{\dagger}{\cal U},$
which give the conditions in coin-space,
\begin{eqnarray}
{\cal I}_{d} & = & A^{\dagger}A+B^{\dagger}B+M^{\dagger}M,\nonumber\\
0&=&A^{\dagger}M+M^{\dagger}B=A^{\dagger}B,\label{eq:unitarityCondition}
\end{eqnarray}
implying that $A+B+M$ is unitary. As ${\cal C}$ is unitary,
these conditions equally apply to $P$, $Q$, and $R$. They can not
be satisfies by scalars (except for trivial cases).\cite{Meyer96,Portugal14}

The algebra in Eq.~(\ref{eq:unitarityCondition}) requires at least
two-dimensional matrices, and matching the dimension $c$ of the coin
space and the degree of each site represents a natural and commonly
studied choice. For the $d$-dimensional hypercubic lattice, this
means $c=2d$, but higher dimensional coins\cite{Inui05,Falkner14a}
and even coinless alternatives\cite{Meyer96,Patel05,Fal13,Ambainis13,Portugal14} have been studied.

\begin{figure}[t]
\includegraphics[bb=0bp 0bp 792bp 450bp,clip,width=1\columnwidth]{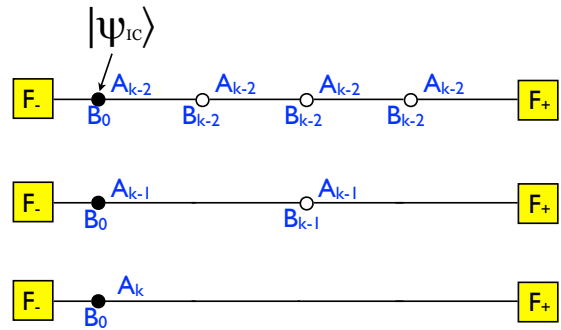}
\caption{\label{fig:1dQWRG}(Color Online) Absorption model for a simple line, indicating
the three final recursion steps. Boxes represent absorbing sites,
black-marked sites the initial conditions $\left|\psi_{IC}\right\rangle $.
Labels $A$ and $B$ indicate the respective hopping parameters for
each site.}
\end{figure}

\section{Renormalization of the Quantum Walk on a Line\label{sec:Renormalization-1dQW}}

We introduce generating functions,\cite{Redner01,SWlong}
\begin{equation}
\tilde{\psi}_{x}\left(z\right)={\textstyle \sum_{t=0}^{\infty}}\psi_{x,t}z^{t},\label{eq:LaplaceT}
\end{equation}
to eliminate the explicit time-dependence, which allows us to obtain the RG-recursions. The asymptotic behavior for $t\to\infty$ is obtained in the limit of $z\to1$, which puts more weight on terms with high values of $t$ in Eq.~(\ref{eq:LaplaceT}). In the inverse-transform of Eq.~(\ref{eq:LaplaceT}), the limit $z\to1$ is intimately related to the large-time limit due to the cross-over at $t(1-z)\sim1$ in $z^t=\exp\{t\ln z\}\sim\exp\{-t(1-z)\}$, see Ref.~\cite{Redner01} or any textbook on generating functions.

The master equation (\ref{eq:MasterE}) with ${\cal U}$ in Eq.~(\ref{eq:propagator-1})
then becomes
\begin{equation}
\tilde{\psi}_{x}=zM\tilde{\psi}_{x}+zA\tilde{\psi}_{x-1}+zB\tilde{\psi}_{x+1}+\delta_{x,0}\psi_{IC}.
\end{equation}
For simplicity, we merely consider initial conditions (IC) localized
at the origin, $\psi_{x,t=0}=\delta_{x,0}\psi_{IC}$. As depicted
in Fig.~\ref{fig:1dQWRG}, we recursively eliminate $\tilde{\psi}_{x}$
for all sites for which $x$ is an odd number, then set $x\to x/2$
and repeat, step-by-step for $k=0,1,2,\ldots$. Each such step corresponds
to a rescaling of the system size by a factor of 2, and after $k$
iterations, $\tilde{\psi}_{x}^{(k)}$ represents the renormalized
wave-function describing a domain of size $2^{k}$, and the corresponding
renormalized hopping parameters describe the effective transport in
and out of that domain. 

To wit, starting at $k=0$ with the ``raw'' hopping coefficients
$A_{0}=zA$, $B_{0}=zB$, and $M_{0}=zM$, after each step, the master
equation becomes self-similar in form when redefining the renormalized
hopping coefficients $A_{k}$, $B_{k}$, $M_{k}$. For example, for
consecutive sites near any even site $x$ at step $k$ we have:\cite{Boettcher13a}
\begin{eqnarray}
\tilde{\psi}_{x-1} & = & M_{k}\tilde{\psi}_{x-1}+A_{k}\tilde{\psi}_{x-2}+B_{k}\tilde{\psi}_{x},\nonumber \\
\tilde{\psi}_{x} & = & M_{k}\tilde{\psi}_{x}+A_{k}\tilde{\psi}_{x-1}+B_{k}\tilde{\psi}_{x+1}+\delta_{x,0}\psi_{IC},\label{eq:1dPRWmass-master}\\
\tilde{\psi}_{x+1} & = & M_{k}\tilde{\psi}_{x+1}+A_{k}\tilde{\psi}_{x}+B_{k}\tilde{\psi}_{x+2}.\nonumber 
\end{eqnarray}
Solving for the central site $x$ yields
\begin{equation}
\tilde{\psi}_{x}=M_{k+1}\tilde{\psi}_{x}+A_{k+1}\tilde{\psi}_{x-2}+B_{k+1}\tilde{\psi}_{x+2}+\delta_{x,0}\psi_{IC},
\end{equation}
with RG ``flow'' 
\begin{eqnarray}
A_{k+1} & = & A_{k}\left(I-M_{k}\right)^{-1}A_{k},\nonumber \\
B_{k+1} & = & B_{k}\left(I-M_{k}\right)^{-1}B_{k},\label{eq:recur1dPRWmass}\\
M_{k+1} & = & M_{k}+A_{k}\left(I-M_{k}\right)^{-1}B_{k}+B_{k}\left(I-M_{k}\right)^{-1}A_{k},\nonumber 
\end{eqnarray}
where the hopping parameters in general are matrices.

\subsubsection{Example: RG for the Classical Random Walk\label{sub:Example:-RG-for}}

In the classical analysis\cite{Redner01,SWlong} for a random walk
with a Bernoulli coin $p$, Eqs.~(\ref{eq:recur1dPRWmass}) reduce
to
\begin{align}
a_{k+1} & =\frac{a_{k}^{2}}{1-m_{k}},\nonumber \\
b_{k+1} & =\frac{b_{k}^{2}}{1-m_{k}},\label{eq:1dRWmass}\\
m_{k+1} & =m_{k}+\frac{2a_{k}b_{k}}{1-m_{k}},\nonumber
\end{align}
with \emph{scalar} quantities, which initiate at $k=0$ with $a_{0}=zp$,
$b_{0}=z\left(1-p\right)$, and $m_{0}=0$. The fixed points (FP)
arising from this RG-flow for $k\sim k+1\to\infty$ at $z\to1$ are
$\left(a_{\infty},b_{\infty},m_{\infty}\right)=(0,0,m_{\infty})$,
$\left(1-m_{\infty},0,m_{\infty}\right)$, or $\left(0,1-m_{\infty},m_{\infty}\right)$
for any value of $m^{*}$ on the unit interval.%
\footnote{ From the exact solution one finds that in this case the FP value
$m^{*}$ depends in the initial values for the RG-flow, $1-m^{*}=\left|a_{0}-b_{0}\right|_{z=1}$,
which can not be obtained from the asymptotic analysis.}

Perturbing the RG-flow in  Eqs.~(\ref{eq:1dRWmass}) via $\{a,b,m\}_k\sim \{a,b,m\}_\infty+(1-z)\{\alpha,\beta,\mu\}_k$  for $z\to1$ and large $k$, we find the linear system $(\alpha,\beta,\mu)_{k+1}=J\circ(\alpha,\beta,\mu)_k$ with the Jacobian
\begin{eqnarray}
J & = & \left.\frac{\partial\left(a_{k+1},b_{k+1},m_{k+1}\right)}{\partial\left(a_{k},b_{k},m_{k}\right)}\right|_{k\to\infty},\label{eq:Jacobian}\\
 & = & \left(\begin{array}{ccc}
\frac{2a_{\infty}}{1-m_{\infty}}, & 0, & \frac{2b_{\infty}}{1-m_{\infty}}\\
0, & \frac{2b_{\infty}}{1-m_{\infty}}, & \frac{2a_{\infty}}{1-m_{\infty}}\\
\frac{a_{\infty}^{2}}{\left(1-m_{\infty}\right)^{2}}, & \frac{b_{\infty}^{2}}{\left(1-m_{\infty}\right)^{2}}, & 1+\frac{2a_{\infty}b_{\infty}}{\left(1-m_{\infty}\right)^{2}}
\end{array}\right).\nonumber 
\end{eqnarray}
The largest eigenvalue $\lambda_w$ of this Jacobian, via $t(1-z)\sim1$, then describes how time 
rescales, $t_{k}=\lambda_{w}t_{k-1}$, when doubling system length, $L_{k}=2L_{k-1}$.  Assuming a similarity  solution for the probability density function of the walk, $\rho\left(x,t\right)\sim f\left(x^{d_{w}}/t\right)$, the scaling Ansatz relating distance and time, $t_{k}\sim L_{k}^{d_{w}}$, thus provides 
\begin{equation}
d_{w}=\log_{2}\lambda_{w}. 
\label{eq:dw}
\end{equation}

Inserting the \emph{2nd} and \emph{3rd} FP in $J$ easily yield the ballistic solutions, $d_w=1$, for drifting either to the left or to the right only. In contrast,
the indeterminedness of $m_{\infty}$ in the 1st FP is peculiar. In
fact, for $z=1$, $a_{k}+b_{k}+m_{k}=1$ for all $k$ and
starting from symmetric initial values $a_{0}=b_{0}$, i.e., $p=\frac{1}{2}$,
both remain identical and vanish together, $a_{k}\equiv b_{k}\to0$,
and $m_{k}\to m_{\infty}=1$. Since both numerators and denominators
in the Jacobian vanish, a correlated solution has to be constructed
that ``peals off'' the leading behavior to glance into the boundary
layer. Using $a_{k}\equiv b_{k}\sim a_{k}^{\prime}\epsilon^{k}$ and
$m_{k}\sim1-m_{k}^{\prime}\epsilon^{k}$ assuming large $k$ and $\left|\epsilon\right|<1$,
results in
\begin{equation}
a_{k+1}^{\prime}=\frac{a_{k}^{\prime2}}{\epsilon m_{k}^{\prime}},\qquad m_{k+1}^{\prime}=\frac{1}{\epsilon}m_{k}^{\prime}-\frac{2a_{k}^{\prime2}}{\epsilon m_{k}^{\prime}}
\end{equation}
with a single FP that self-consistently determines $\frac{a_{\infty}^{\prime}}{m_{\infty}^{\prime}}=\epsilon=\frac{1}{2}$.
The Jacobian of these recursions, $J^{\prime}=\frac{\partial\left(a_{k+1}^{\prime},m_{k+1}^{\prime}\right)}{\partial\left(a_{k}^{\prime},m_{k}^{\prime}\right)}|_{k\to\infty}$,
at its FP gives $\lambda_{w}=4$ as the largest eigenvalue, i.e.,
$d_{w}=2$ for the diffusive solution. In this formulation, even if
we start with vanishing self-term initially, $m_{0}=0$, the self-term
ultimately dominates, $m_k\to1$, reflecting the fact that in diffusion the renormalized
domain of length $L_{k}\sim2^{k}$ outgrows the walk such that almost all
hops remain within that domain. 

We note that the RG projects the salient properties of the walk
into three ``universality classes:'' diffusive and ballistic motion
either to the left or right, characterized by an exponent $d_{w}=2$
or 1. Each characterizes a fixed point of the dynamics, reached either
for $p=\frac{1}{2}$ or $p\not=\frac{1}{2}$.

\subsubsection{RG for the Quantum Walk on a Line\label{sub:RG-for-1dQW}}

For the quantum walk, we set 
\begin{equation}
A_{k}=P_{k}{\cal C},\qquad B_{k}=Q_{k}{\cal C},\qquad M_{k}=R_{k}{\cal C},\label{eq:PQRtoABM-QW}
\end{equation}
where initially $P_{0}=zP$, $Q_{0}=zQ$, and $R_{0}=zR$ from Eq.
(\ref{eq:PQRinit-1dQW}). To gain an intuition, we evolve the RG-flow
(\ref{eq:recur1dPRWmass}) for a few iterations from these raw values.
Each iteration consists of assembling the hopping parameters at level
$k$ according to Eq.~(\ref{eq:PQRtoABM-QW}), the actual RG-step
of applying Eqs.~(\ref{eq:recur1dPRWmass}), and then of \emph{inverting}
Eq.~(\ref{eq:PQRtoABM-QW}) with ${\cal C}^{-1}$ to arrive at $P_{k+1}$,
$Q_{k+1}$, and $R_{k+1}$. Already after two steps, a recurring pattern
emerges that suggests the parametrization
\begin{equation}
P_{k}=\left(\begin{array}{cc}
a_{k} & 0\\
0 & 0
\end{array}\right),\enskip Q_{k}=\left(\begin{array}{cc}
0 & 0\\
0 & -a_{k}
\end{array}\right),\enskip R_{k}=\left(\begin{array}{cc}
0 & m_{k}\\
m_{k} & 0
\end{array}\right).\label{eq:1dPRWAnsatz}
\end{equation}
 Indeed, for $a_{k}$ and $m_{k}$, the RG-flow (\ref{eq:recur1dPRWmass})
closes after each iteration with 
\begin{eqnarray}
a_{k+1} & = & \frac{a_{k}^{2}\sin\eta}{1-2m_{k}\cos\eta+m_{k}^{2}},\nonumber \\
m_{k+1} & = & m_{k}+\frac{\left(m_{k}-\cos\eta\right)a_{k}^{2}}{1-2m_{k}\cos\eta+m_{k}^{2}}\label{eq:1dQWrecursions1}
\end{eqnarray}
for $0<\eta<\pi/2$ (setting $\chi=\vartheta=0$). These recursions
have a single fixed point at $\left(a_{\infty},m_{\infty}\right)=\left(\sin\eta,\cos\eta\right)$,
yet, its Jacobian at $k\to\infty$ is $\eta$-independent and has
a degenerate eigenvalue $\lambda_{w}=2$, suggesting $d_{w}=\log_{2}\lambda=1$.
This reflects the well-known universality of the large-scale dynamics
of the quantum walk on the line with respect to the chosen coin.\cite{Bach2004562} 

\begin{figure}[b]
\includegraphics[bb=0bp 80bp 792bp 550bp,clip,width=1\columnwidth]{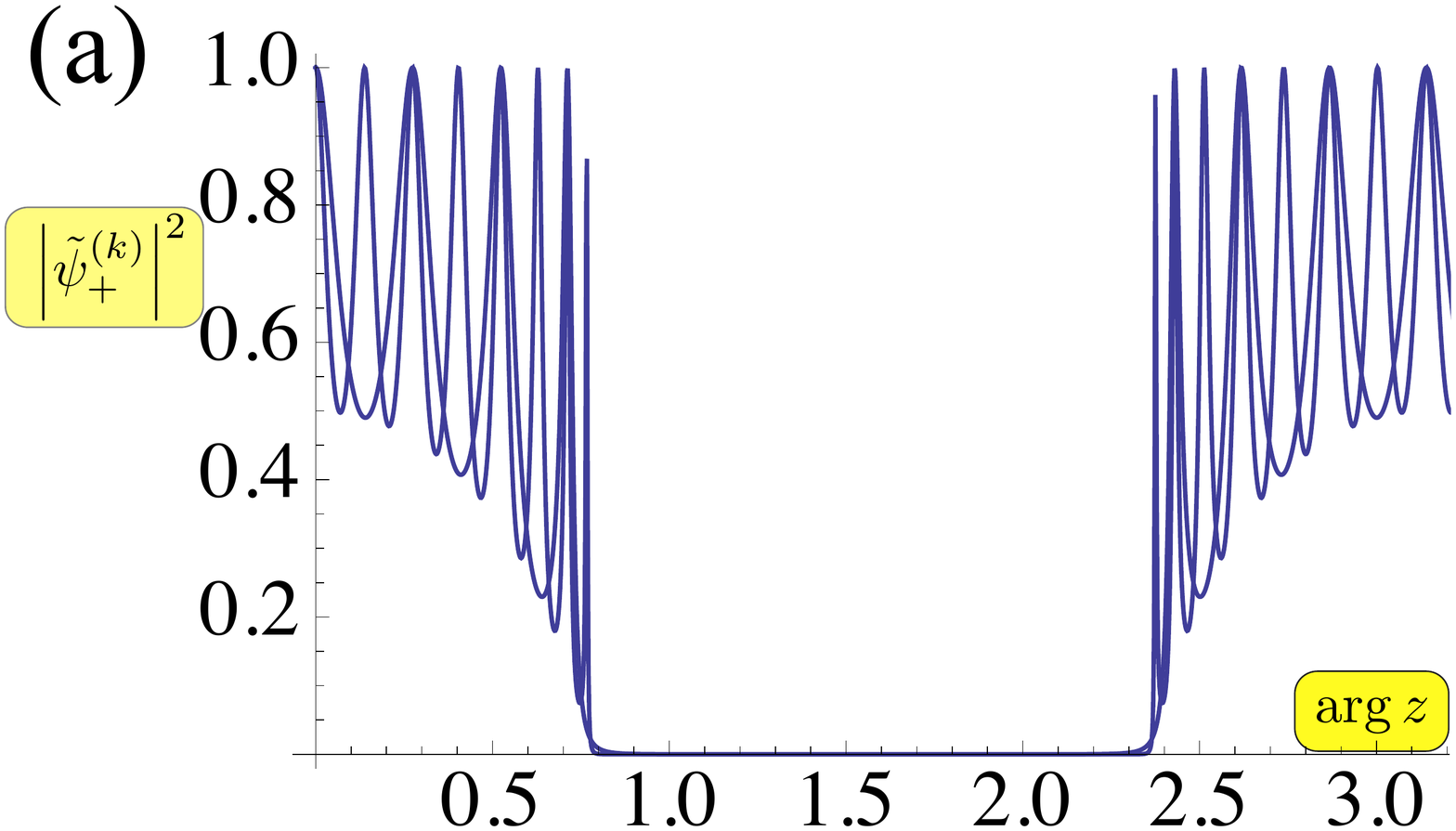}

\includegraphics[bb=0bp 80bp 792bp 550bp,clip,width=1\columnwidth]{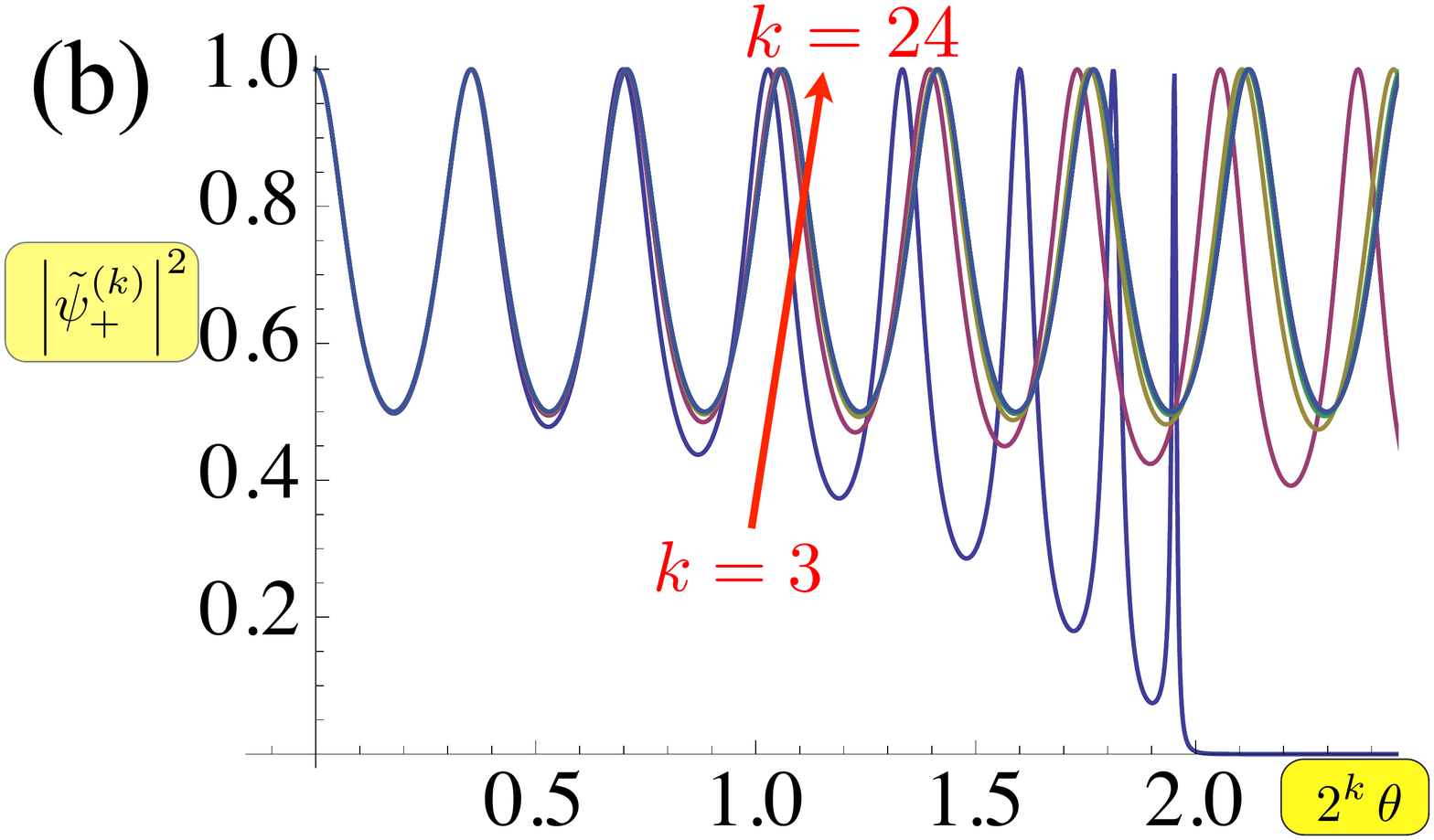}
\caption{\label{fig:Fpm1dQW}(Color Online)  Plot of $\left|\tilde{\psi}_{+}^{\left(k\right)}\left(z\right)\right|$
at $\eta=\pi/4$ and generic initial conditions $\psi_{IC}=\left(0,1\right)$
(a) for $\arg z$ after $k=2,3$ iterations and (b) for the rescaled
variable $2^{k}\theta$, $z=-e^{i\theta}$, to collapse all data up
to $k=24$ for $\theta\to0$. As (a) shows, the integrand is periodic
with period $\pi$ and has significant support only for $|\theta|<\eta$
and $\pi-|\theta|<\eta$. }
\end{figure}

As we will show in Sec.~\ref{sec:RG-Fixed-Points-are}, though, this
picture may be incomplete, a lucky accident due to the fact that the ``fractal''
exponent, $d_{f}=\log_{2}\lambda_{f}=1$, and the walk exponent $d_{w}$
coincide. To exemplify this aspect here, let us consider the probability
$F_{a}$ of \emph{ever} being absorbed at a site $x=a$ as a simple and generic
observable.\cite{Ambainis01,Bach2004562} From Eq.~(\ref{eq:LaplaceT}),
for a random walk with $\rho\left(x,t\right)=\psi_{x,t}$, it is simply
$F_{a}=\sum_{t}\psi_{a,t}=\lim_{z\to1}\tilde{\psi}_{a}\left(z\right)$.
For a quantum walk, it is instead $\rho\left(x,t\right)=\left|\psi_{x,t}\right|^{2}$,
and hence, 
\begin{equation}
F_{a}=\sum_{t=0}^{\infty}\left|\psi_{a,t}\right|^{2}=\oint\frac{dz}{2\pi iz}\,\left|\tilde{\psi}_{a}\left(z\right)\right|^{2}=\int_{-\pi}^{\pi}\frac{d\theta}{2\pi}\,\left|\tilde{\psi}_{a}\left(\theta\right)\right|^{2},\label{eq:AbsoProb}
\end{equation}
where we choose
\begin{equation}
z=-e^{i\theta}\qquad\left(-\pi<\theta\leq\pi\right),\label{eq:ztheta}
\end{equation}
i.e., $z\to-1$ for $\theta\to0$ as reference point (see below).
While the random walk merely entails a local analysis for real $z\to1$,\cite{Redner01}
the unitarity of quantum walks generally demands an analysis along
the \emph{entire} unit circle in the complex-$z$ plane. Let us put
the quantum walk on the line between two absorbing walls, $F_{\pm}$,
as shown in Fig.~\ref{fig:1dQWRG}, with $F_{-}^{(k)}$ at $x_{-}=-1$
right next to the starting site $x=0$, from which the wall $F_{+}^{(k)}$
at site $x_{+}=2^{k}$ recedes further away with every iteration of
the flow equations. In Fig.~\ref{fig:Fpm1dQW}(a), we plot the integrand
$\left|\tilde{\psi}_{+}^{(k)}\left(z\right)\right|$ for $\arg z=\pi-\theta$,
on the unit circle. Some algebra shows that $\tilde{\psi}_{+}^{(k)}\left(z\right)\propto a_{k}$,
which depends on $z$ through $a_{0}$. The asymptotic behavior of
$a_{k}$ in Eq.~(\ref{eq:1dQWrecursions1}) for large $k$ at fixed
$\theta$ falls into one of four different cases: (i) at $\theta=0$
and $\theta=\pi$, the stationary behavior for the aforementioned
fixed point is obtained; (ii) for $0<\left|\theta\right|<\eta$ and
$0<\pi-\left|\theta\right|<\eta$, $a_{k}$ varies \emph{chaotically}
with $k$; (iii) for $\left|\theta\right|=\eta$ and $\left|\theta\right|=\pi-\eta$
local analysis recovers classical diffusive scaling; and (iv) for
$\eta<\left|\theta\right|<\pi-\eta$, $a_{k}$ vanishes exponentially
with $N=2^{k}$. We argue that only the chaotic regimes, $\left|\theta\right|<\eta$
and $\pi-\left|\theta\right|<\eta$, with the stationary point at
their center contributes to extensive quantum transport.%
\footnote{Non-zero values for $\chi,\vartheta$ in Eq.~(\ref{eq:QW1dCoin-1})
merely rotate this picture around the unit circle, i.e., affect a
trivial shift in $\theta$.%
} To wit, it can be shown that for $\eta\to0$ the ``velocity'' of
the ballistically spreading quantum walk decreases to zero
and it eventually becomes localized for $\eta=0$, exactly when both
chaotic regimes shrink to zero, while the stationary point remains
inside. 

Clearly, quantum transport here is determined by properties of the
entire wave-function, $\tilde{\psi}_{x}\left(z\right)$, not just
the limit $z\to z_{0}$ near some fixed point $z_{0}$. We will analyze
this situation in more detail in Sec.~\ref{sec:RG-Fixed-Points-are}.

\begin{figure}
\includegraphics[bb=0bp 0bp 700bp 612bp,clip,width=0.7\columnwidth]{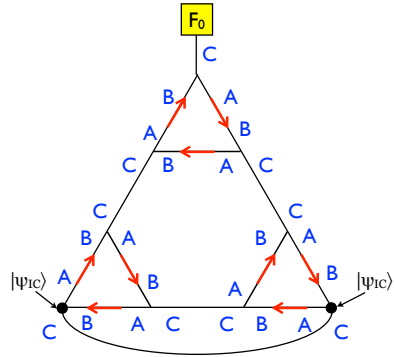}

\caption{\label{fig:Absorption-DSG}(Color Online) Absorption model for the dual Sierpinski
gasket at generation $k=1$. Boxes represent absorbing site, black-marked sites harbor
the initial conditions $\left|\psi_{IC}\right\rangle $. Labels $A,B,C$
indicate the respective hopping parameters for leaving each site. Note that for increasing generation $k$, the minimal separation between initiating sites and the sole absorbing site will increase as $\sim2^k$.}

\end{figure}

\section{Renormalization of the Quantum Walk on the Dual Sierpinski Gasket\label{sec:Renormalization-DSG}}

The DSG is a degree-3 lattice, see Fig.~\ref{fig:Absorption-DSG},
and unitarity requires at least a $c=3$-dimensional coin. As the
most general such coin has six real parameters, we focuses here only
on the real and symmetric Grover coin, 
\begin{equation}
{\cal C}=\frac{1}{3}\left(\begin{array}{ccc}
-1 & 2 & 2\\
2 & -1 & 2\\
2 & 2 & -1
\end{array}\right),\label{eq:3x3Grover}
\end{equation}
and defer generalizations to future discussions. Note that DSG has
several advantages over the more familiar Sierpinski gasket (of which
it is dual). DSG is of regular degree 3, while Sierpinski itself is
a regular degree-4 lattice. However, it is not merely the higher degree
that provides difficulties for quantum walks on the original Sierpinski
gasket. The internal coin degrees of freedom provide a labeling problem
that severely complicates its consideration, already in evidence in
Ref.~\onlinecite{Lara12}. Interestingly, none of these problems exist for
the classical random walk, and the Sierpinski gasket (or its dual)
serves as a popular example of a simple demonstration of the RG, as
its hierarchical structure and high degree of symmetry affects an
RG-flow in a \emph{single} real hopping parameter. For the quantum
walk, we will find instead \emph{five} coupled complex recursions
with a large number of terms.

\begin{figure*}
\includegraphics[clip,width=0.65\paperwidth]{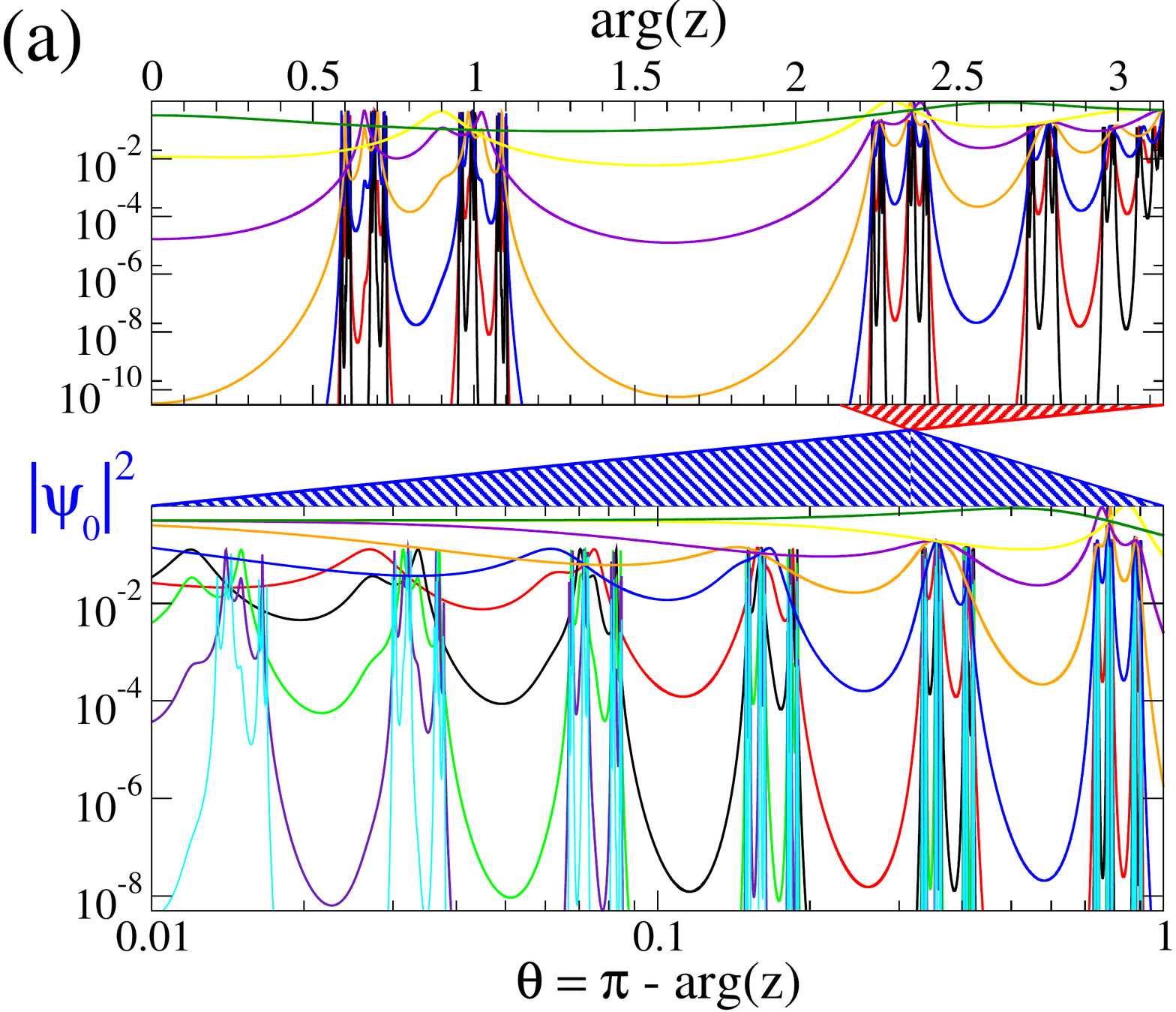}

\includegraphics[bb=0bp 70bp 792bp 612bp,clip,width=0.65\paperwidth]{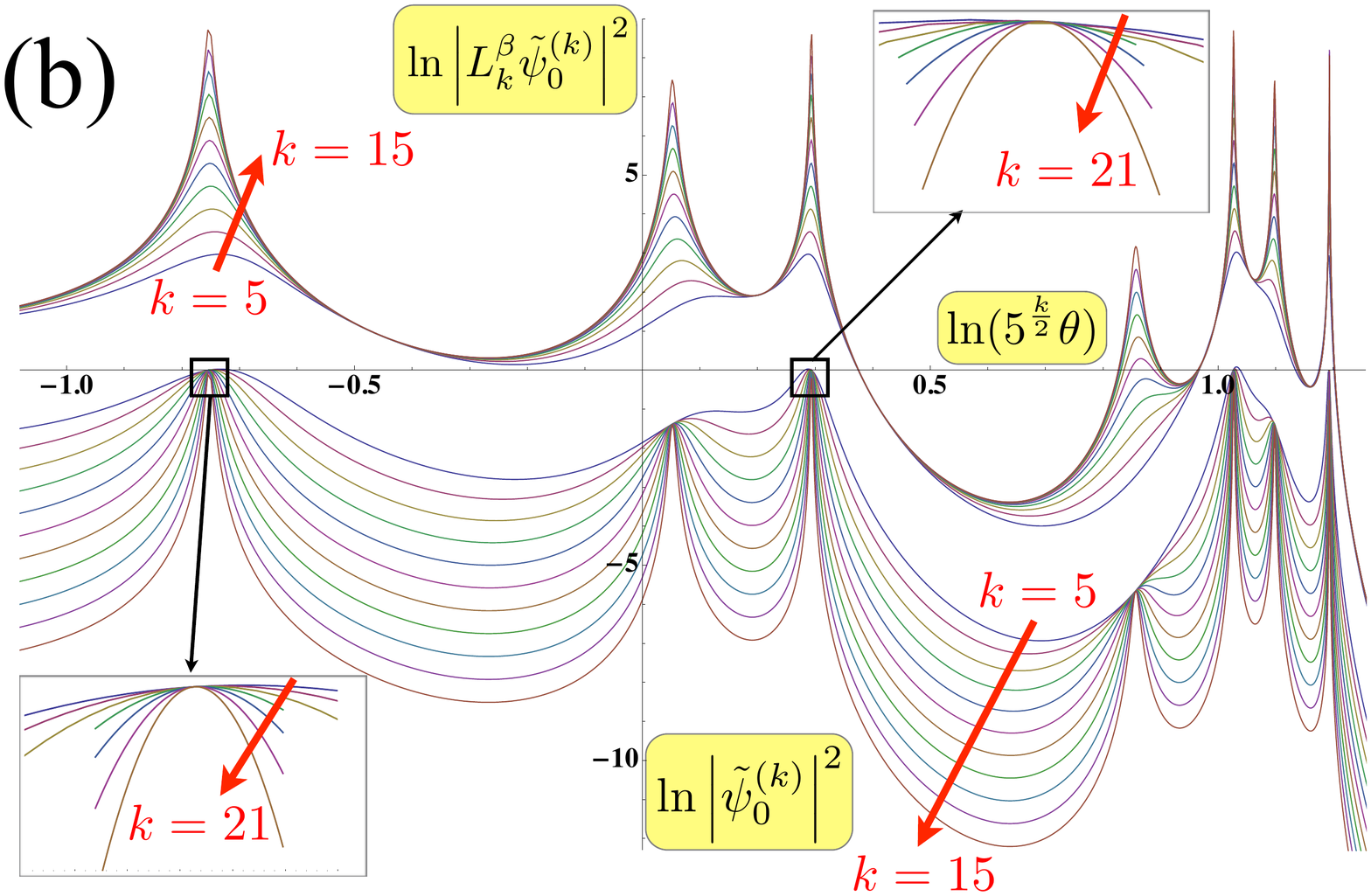}
\caption{\label{fig:collapseDSG}(Color Online) (a) Log-plot of the integrand $\left|\tilde{\psi}_{0}^{\left(k\right)}\left(z\right)\right|$
in Eq.~(\ref{eq:AbsoProb}) for $0\leq\arg z\leq\pi$ for $k=1,\ldots6$.
Decaying almost everywhere, the integrand has most support near the
fixed point, $\arg z\to\pi$. A logarithmic scale with $\theta=\pi-\arg z\to0$
reveals a self-similar sequence of periodic structures. (b) Scaling
collapse of $\left|L_{k}^{\beta}\tilde{\psi}_{0}^{\left(k\right)}\left(z\right)\right|$
with $L_{k}=2^{k}$ for $\ln\left(\lambda_{w}^{k}\theta\right)$ at
$k=5,\ldots,15$ using $\lambda_{w}=\sqrt{5}$. At $\beta=0$, all
data lines up, in particular, all peaks remain constant (=1 up to
$k=21$, see insets) but narrow. At $\beta=0.424(3),$ all data in
the fast-decaying intervals collapse but the peaks now diverge. }
\end{figure*}

\subsection{Renormalization of the Dual Sierpinski Gasket\label{sec:RG-Flow-for-DSG}}

Fig.~\ref{fig:Absorption-DSG} shows the elementary graph-let of
nine sites that is used to construct the dual Sierpinski gasket (DSG).
It also represents the basic unit from which we can extract the flow
equations by tracing out the wave functions $\tilde{\psi}_{4},\ldots,\tilde{\psi}_{9}$
on the six inner sites to leave only the three corner sites $\tilde{\psi}_{1},\ldots,\tilde{\psi}_{3}$
and the renormalized hopping coefficients between them. Note the systematic
labeling for the three out-direction of each site that determines,
in effect, which shift operator applies to that direction. The master
equations relating those nine sites are: 
\begin{eqnarray}
\tilde{\psi}_{1} & = & M_{k}\tilde{\psi}_{1}+C_{k}\tilde{\psi}_{\bar{1}}+A_{k}\tilde{\psi}_{4}+B_{k}\tilde{\psi}_{5},\nonumber \\
\tilde{\psi}_{2} & = & M_{k}\tilde{\psi}_{2}+C_{k}\tilde{\psi}_{\bar{2}}+A_{k}\tilde{\psi}_{6}+B_{k}\tilde{\psi}_{7},\nonumber \\
\tilde{\psi}_{3} & = & M_{k}\tilde{\psi}_{3}+C_{k}\tilde{\psi}_{\bar{3}}+A_{k}\tilde{\psi}_{8}+B_{k}\tilde{\psi}_{9},\nonumber \\
\tilde{\psi}_{4} & = & M_{k}\tilde{\psi}_{4}+C_{k}\tilde{\psi}_{9}+A_{k}\tilde{\psi}_{5}+B_{k}\tilde{\psi}_{1},\nonumber \\
\tilde{\psi}_{5} & = & M_{k}\tilde{\psi}_{5}+C_{k}\tilde{\psi}_{6}+A_{k}\tilde{\psi}_{1}+B_{k}\tilde{\psi}_{4},\label{eq:PRWmasterDSGmass}\\
\tilde{\psi}_{6} & = & M_{k}\tilde{\psi}_{6}+C_{k}\tilde{\psi}_{5}+A_{k}\tilde{\psi}_{7}+B_{k}\tilde{\psi}_{2},\nonumber \\
\tilde{\psi}_{7} & = & M_{k}\tilde{\psi}_{7}+C_{k}\tilde{\psi}_{8}+A_{k}\tilde{\psi}_{2}+B_{k}\tilde{\psi}_{6},\nonumber \\
\tilde{\psi}_{8} & = & M_{k}\tilde{\psi}_{8}+C_{k}\tilde{\psi}_{7}+A_{k}\tilde{\psi}_{9}+B_{k}\tilde{\psi}_{3},\nonumber \\
\tilde{\psi}_{9} & = & M_{k}\tilde{\psi}_{9}+C_{k}\tilde{\psi}_{4}+A_{k}\tilde{\psi}_{3}+B_{k}\tilde{\psi}_{8}.\nonumber 
\end{eqnarray}
Here, $\tilde{\psi}_{\bar{1}},\ldots,\tilde{\psi}_{\bar{3}}$ refer
to the corner sites of the respective neighboring graph-lets, which
themselves also do not get renormalized.

It drastically reduces the algebraic effort to trace out $\tilde{\psi}_{4},\ldots,\tilde{\psi}_{9}$
in a symmetrical way. When eliminated, each of those amplitudes must
be a function of the remaining three, $\tilde{\psi}_{1},\ldots,\tilde{\psi}_{3}$,
in a cyclically permuted manner. Thus, we start with the Ansatz 
\begin{eqnarray}
\tilde{\psi}_{4} & = & a\tilde{\psi}_{1}+b\tilde{\psi}_{2}+c\tilde{\psi}_{3},\nonumber \\
\tilde{\psi}_{5} & = & d\tilde{\psi}_{1}+e\tilde{\psi}_{2}+f\tilde{\psi}_{3},\label{eq:AlgebraAnsatzDSG}
\end{eqnarray}
and similarly for the inner sites at there other two corners, $\left(\tilde{\psi}_{6},\tilde{\psi}_{7}\right)$
and $\left(\tilde{\psi}_{8},\tilde{\psi}_{9}\right)$, by appropriately
permuting the indices on $\tilde{\psi}_{1},\ldots,\tilde{\psi}_{3}$.
Inserting these prospective solutions into the right-hand side of
the last six relations in Eqs.~(\ref{eq:PRWmasterDSGmass}) and comparing
coefficients with Eqs.~(\ref{eq:AlgebraAnsatzDSG}) provides self-consistency
relations for the matrices $a,\ldots,f$. This step eliminates $\tilde{\psi}_{4},\ldots,\tilde{\psi}_{9}$
by transforming the problem into one of expressing the matrices $a,\ldots,f$
in terms of $A_{k},B_{k},C_{k},$ and $M_{k}$, or with simpler notation:
$\bar{A}=\left(I-M_{k}\right)^{-1}A_{k}$, $\bar{B}=\left(I-M_{k}\right)^{-1}B_{k}$,
and $\bar{C}=\left(I-M_{k}\right)^{-1}C_{k}$. Most importantly, this
Ansatz has disentangled the original six equations into two \emph{equivalent,
closed} sets of three relations that can be solved independently:
Since comparing coefficients provides a bipartite set of relations
initially,
\begin{equation}
\begin{array}{rclcrcl}
a & = & \bar{A}d+\bar{C}e+\bar{B}, & \quad & d & = & \bar{B}a+\bar{C}c+\bar{A},\\
b & = & \bar{A}e+\bar{C}f, & \quad & e & = & \bar{B}b+\bar{C}a,\\
c & = & \bar{A}f+\bar{C}d, & \quad & f & = & \bar{B}c+\bar{C}b,
\end{array}
\end{equation}
 (noting that these are non-commuting matrices), we write 
\begin{eqnarray}
a & = & \bar{A}\bar{B}a+\bar{A}\bar{C}c+\bar{A}^{2}+\bar{C}\bar{B}b+\bar{C}^{2}a+\bar{B},\nonumber\\
b & = & \bar{A}\bar{B}b+\bar{A}\bar{C}a+\bar{C}\bar{B}c+\bar{C}^{2}b,\\
c & = & \bar{A}\bar{B}c+\bar{A}\bar{C}b+\bar{C}\bar{B}a+\bar{C}^{2}c+\bar{C}\bar{A},\nonumber
\end{eqnarray}
and an corresponding set for $d,e,f$ by identifying $a\Leftrightarrow d$,
$b\Leftrightarrow f$, $c\Leftrightarrow e$, and $\bar{A}\Leftrightarrow\bar{B}$,
while $\bar{C}$ remains in place. To minimize the number of matrix-multiplications,
it is now convenient to abbreviate 
\begin{eqnarray}
V & = & \left(I-\bar{A}\bar{B}-\bar{C}^{2}\right)^{-1},\nonumber\\
{\cal A} & = & V\bar{A}\bar{C},\nonumber\\
{\cal B} & = & V\bar{C}\bar{B},\nonumber\\
W & = & \left(I-{\cal B}{\cal A}\right)^{-1},\\
X & = & {\cal A}^{2}+{\cal B},\nonumber\\
Y & = & W\left({\cal A}+{\cal B}^{2}\right),\nonumber\\
Z & = & WV\left(A^{2}+B\right).\nonumber
\end{eqnarray}
With those, we find 
\begin{eqnarray}
c & = & \left[I-{\cal A}{\cal B}-XY\right]^{-1}\left(V\bar{C}\bar{A}+XZ\right),\nonumber\\
a & = & Yc+Z,\\
b & = & {\cal A}a+{\cal B}c,\nonumber
\end{eqnarray}
and the complementing set for $d,e,$ and $f$. Finally, inserting
$\tilde{\psi}_{4},\tilde{\psi}_{5}$ from Eqs.~(\ref{eq:AlgebraAnsatzDSG})
into the relation for $\tilde{\psi}_{1}$ (or, respectively, inserting
$\tilde{\psi}_{6},\tilde{\psi}_{7}$ into the relation for $\tilde{\psi}_{2}$,
or $\tilde{\psi}_{8},\tilde{\psi}_{9}$ for $\tilde{\psi}_{3}$) yields
the renormalization-flow 
\begin{eqnarray}
M_{k+1} & = & M_{k}+A_{k}a+B_{k}d,\nonumber \\
A_{k+1} & = & A_{k}c+B_{k}f,\nonumber \\
B_{k+1} & = & A_{k}b+B_{k}e,\label{eq:RGflowDSGPRW}\\
C_{k+1} & = & C_{k}.\nonumber 
\end{eqnarray}
(Note that $C_{k}$ does not renormalize.) While tedious to derive,
these equations are exact and easily implemented on a computer algebra
system.

\subsection{Parametrizing the RG-Flow for the Dual Sierpinski Gasket}

Similar to Sec.~\ref{sub:RG-for-1dQW}, we define shift matrices 
\begin{equation}
P=\left[\begin{array}{ccc}
1 & 0 & 0\\
0 & 0 & 0\\
0 & 0 & 0
\end{array}\right],\quad Q=\left[\begin{array}{ccc}
0 & 0 & 0\\
0 & 1 & 0\\
0 & 0 & 0
\end{array}\right],\quad R=0.
\end{equation}
Since $C_{k}=C_{0}$, it will require no parametrization. $ $We initiate
the recursions at $k=0$ with $A_{0}=zPC$, $B_{0}=zQ{\cal C}$, and
$M_{0}=zR{\cal C}=0$. After a single iteration, a recursive pattern
emerges that suggests a 5-parameter Ansatz: 
\begin{eqnarray}
P_{k} & = & \left(\begin{array}{ccc}
a_{k}^{(1)} & a_{k}^{(2)} & 0\\
a_{k}^{(2)} & a_{k}^{(3)} & 0\\
0 & 0 & 0
\end{array}\right),\quad Q_{k}=\left(\begin{array}{ccc}
a_{k}^{(3)} & a_{k}^{(2)} & 0\\
a_{k}^{(2)} & a_{k}^{(1)} & 0\\
0 & 0 & 0
\end{array}\right),\nonumber \\
R_{k} & = & \frac{1}{2}\left(\begin{array}{ccc}
m_{k}^{(1)}-m_{k}^{(2)} & m_{k}^{(1)}+m_{k}^{(2)} & 0\\
m_{k}^{(1)}+m_{k}^{(2)} & m_{k}^{(1)}-m_{k}^{(2)} & 0\\
0 & 0 & 0
\end{array}\right),\label{eq:RGbeforePRWDSG}
\end{eqnarray}
Iteration provides of \emph{closed} set of five complex recursions,
$\left\{ a_{k+1}^{(1,2,3)},m_{k+1}^{(1,2)}\right\} ={\cal R}\left(\left\{ a_{k}^{(1,2,3)},m_{k}^{(1,2)}\right\} ;z\right)$,
each a ratio of polynomials similar to Eqs.~(\ref{eq:1dQWrecursions1})
but with \emph{dozens} of terms and an explicit dependence on $z$
due to $C_{0}$. Again, the algebra is easily handled on the computer,
however, these recursions prove numerically unstable, and numerical
precision is quickly lost near points of interest, such as $z=-1$.

\subsection{Asymptotic properties of the RG-flow\label{sub:Asymptotic-properties-of}}

As a specific observable to study, we again focus on the probability
to ever get absorbed at a wall, as shown in Fig.~\ref{fig:Absorption-DSG}.
With increasing length $L$, the sole absorbing site at one corner
of DSG recedes from the starting point of the quantum walk (IC), chosen
at the opposite corners. Thus, the total absorption $F_{0}^{(k)}$
is a measure of quantum transport across the system. Fig.~\ref{fig:collapseDSG}(a)
shows the integrand $\left|\tilde{\psi}_{0}^{\left(k\right)}\left(z\right)\right|^{2}$
of $F_{0}$ in Eq.~(\ref{eq:AbsoProb}) for $0\leq\arg z\leq\pi$
(all observables are symmetric around the real-$z$ axis). We derive
the expression for $\tilde{\psi}_{0}$ in Eq.~(\ref{eq:WallPsiDSG})
in the Appendix. Compared to Fig.~\ref{fig:Fpm1dQW}(a) for the line,
remarkably complex patterns emerge for a quantum walk on DSG: 
\begin{enumerate}
\item There is an \emph{isolated} stationary point at $\arg z=\pi$, i.e.,
$\theta=0$, see Eq.~(\ref{eq:ztheta}), where $a_{k}^{(1,2,3)}=-m_{k}^{(1)}=-\frac{1}{3}$
and $m_{k}^{(2)}=1$ for all $k\geq2$. 
\item A sequence of sparse, rugged peaks that slowly decay seem to accumulate
for $\theta\to0$ with increasing $k$. 
\item Everywhere else the function decays rapidly for increasing $k$, where
$a_{k}^{(1)}\sim a_{k}^{(2)}\sim a_{k}^{(3)}\to0$ and $\left|m_{k}^{(1,2)}\right|\to1$.
There is \emph{no} finite oscillatory domain to signal extensive quantum
transport. Instead, since $\tilde{\psi}_{0}^{\left(k\right)}\sim a_{k}^{(1,2,3)}$,
it is easy to show from the five recursions that for any \emph{fixed}
value of $\theta\not=0$, $\left|\tilde{\psi}_{0}^{\left(k\right)}\left(z\right)\right|\to0$
for $k\to\infty$, suggesting that for large systems quantum transport
ceases such that the absorption approaches zero. 
\end{enumerate}
Unlike for translation-invariant lattices, where some fraction of
quantum walks might localize, on DSG the \emph{entire} walk eventually
gets trapped. However, unlike the sharp localization on lattices,\cite{Inui05,Falkner14a}
the entrapped portion of the wave-function has broad tails here. Only
at $\theta=0$ do we find a fixed point. Its Jacobian has a largest
eigenvalue of $\lambda_{f}=3$, which coincides with the fractal exponent
of DSG, $d_{f}=\log_{2}3$. Yet, the data collapse in Fig.~\ref{fig:collapseDSG}(b)
demonstrates that the limit $\theta\to0$ is singular: all data aligns
and collapses according to Eq.~\eqref{eq:functionalrescale} but
with an eigenvalue  \emph{smaller} than the Jacobian. This collapse
occurs in a regime such that $\lambda_{w}^{k}\theta\sim1$, while
the fixed point at $\ln\theta\to-\infty$ seems infinitely far away
and irrelevant. We will discuss this point in more detail in Sec.
\ref{sec:RG-Fixed-Points-are}.

\begin{figure}[b]
\includegraphics[bb=0bp 20bp 720bp 500bp,width=1\columnwidth]{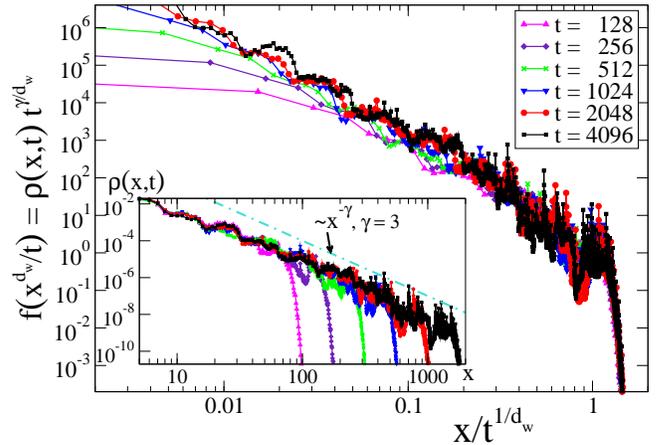}
\caption{\label{fig:NumExtra}(Color Online) Simulations of quantum walks on DSG for the collapse
$f\left(x^{d_{w}}/t\right)\sim\rho\left(x,t\right)t^{\gamma/d_{w}}$
with $d_{w}=\log_{2}\sqrt{5}$ for the probability density function
$\rho\left(x,t\right)=\left|\psi_{x,t}\right|^{2}$ (see inset) for
finding the walker at any site $x$ hops away from the initial sites
at times $t=2^{l}$, $l=7,\ldots,12$, on a DSG of size $N\sim3^{12}$,
before the absorbing wall is reached. To fit not only the cut-off
but also the bulk distribution, we estimate a power-law decay with
exponent $\gamma\approx3$ for $\rho$ as a function of $x$, see
inset.}
\end{figure}

\begin{figure}
\includegraphics[bb=0bp 10bp 720bp 530bp,clip,width=1\columnwidth]{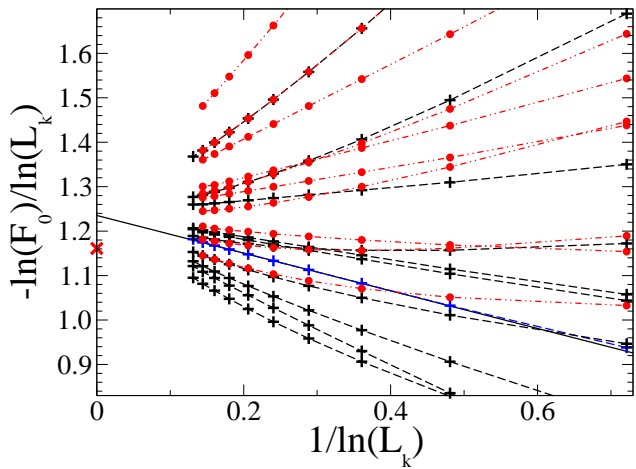}

\caption{\label{fig:Extrapolation}(Color Online) Extrapolation for the scaling exponent in
the decay of the simulated absorption $F_{0}$ with system length
$L_{k}=2^{k}$, $k\leq12$, based on a power-law, for many different
symmetric ($+$) and asymmetric ($\bullet$) initial conditions (IC).
All data appears to extrapolate to the same intercept ($L_{k}\to\infty$)
at about $1.23(1)$, with the most linear fit (extended line) provided
by the symmetric $\psi_{IC}\propto\left(1,2,1\right)$. This exponent
is closely bounded below by $d_{w}=\log_{2}\sqrt{5}$ ($\times$). }
\end{figure}

\begin{figure}
\includegraphics[bb=30bp 20bp 725bp 520bp,clip,width=1\columnwidth]{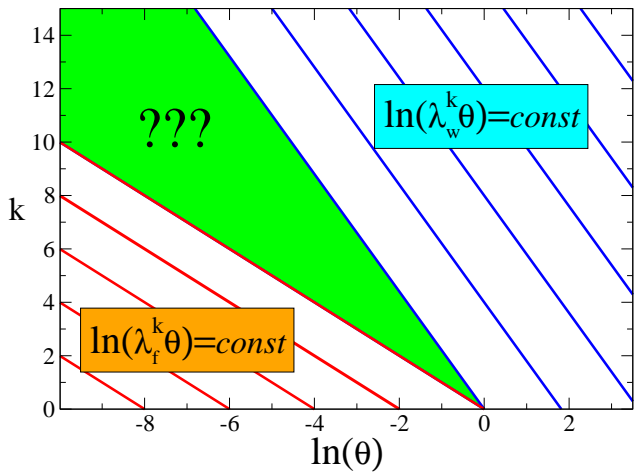}

\caption{\label{fig:RGflowPhase}(Color Online) Depiction of the characteristic trajectories
along which the RG-flow remains invariant, where $\theta=\arg(z-z_{0})$
is a measure of the angular distance from the fixed point at $z_{0}$
in the complex unit circle. In this picture, the fixed point is off
to the left at $\ln\theta=-\infty$, where the data collapses by way
of the fractal exponent $d_{f}=\log_{2}\lambda_{f}$ (red trajectories),
resulting from the largest Jacobian eigenvalue, $\lambda_{f}$. On
the right, where $\lambda_{w}^{k}\theta\gtrsim1$ (blue trajectories),
a far more subtle collapse as that shown in Fig.~\ref{fig:collapseDSG}
results. Note that those trajectories can not ever reach the fixed
point, since $\lambda_{w}<\lambda_{f}$, and will require some intermediate
scaling Ansatz (except for the quantum walk on a line, where $\lambda_{f}=\lambda_{w}$).
The green-shaded area between those different characteristics marks
an unknown cross-over region.}
\end{figure}

The exponents $\lambda_{w}$ and $\beta$ from Eq.~\eqref{eq:functionalrescale}
can be determined recursively with high accuracy from the collapse
with computational cost linear in $k$ (i.e., logarithmic in system
size). As shown at the bottom of Fig.~\ref{fig:collapseDSG}(b) and
especially the insets, lining-up the data for $\left|\tilde{\psi}_{0}^{\left(k\right)}\left(z\right)\right|$,
$k=5,\ldots,21$, without resizing ($\beta=0$) yields $\lambda_{w}=2.23607(2)$,
which we identify as $\lambda_{w}=\sqrt{5}$, such that $d_{w}=\log_{2}\lambda_{w}=\log_{2}\sqrt{5}$.
(The only limitation in the ability to determine the value of $\lambda_{w}$
to arbitrary accuracy numerically is set by the chaotic nature of
the RG-recursions; initiating $z$ near $\theta\to0$ with 1500 digits
accuracy, none remains after only $k=25$ iterations.) 

To demonstrate the relevance of this eigenvalue $\lambda_{w}$ for
the asymptotic spreading of the quantum walk, we have conducted large-scale
numerical simulations directly of the master equation, see Fig.~\ref{fig:NumExtra}.
This value of $d_{w}=\log_{2}\sqrt{5}$ provides data collapse for
$\rho\left(x,t\right)\sim f\left(x^{d_{w}}/t\right)/t^{\gamma/d_{w}}$
with $\gamma\approx3$ from fitting $\rho\left(x,t\right)\sim x^{-\gamma}$,
see inset of Fig.~\ref{fig:NumExtra}. (The origin of this power-law
decay, in contrast to a Gaussian kernel for diffusion, is of yet unknown.)
In particular, the inset shows that the cut-off in $\rho\left(x,t\right)$
scales perfectly as $x_{co}\sim t^{1/d_{w}}$, leading to the collapse
in the main panel. 

Yet, this shift in $\theta$ alone results in a set of functions that \emph{uniformly decay} with
increasing $k$ everywhere but in isolated points, see down-arrows in Fig.~\ref{fig:collapseDSG}(b).
Using $\beta=0.424(3)$ in Eq.~\eqref{eq:functionalrescale} collapses the data everywhere except for
isolated peaks that now \emph{grow} with $k$, see up-arrows in Fig.~\ref{fig:collapseDSG}(b). The
absorption integral from Eq.~(\ref{eq:AbsoProb}) that receives most of its support near
$\left|\theta\right|<\epsilon\ll1$ yields $F_{0}^{(k)}\sim
F_{0}^{(k-1)}/\left(\lambda_{w}2^{2\beta}\right)$, using Eq.~(\ref{eq:functionalrescale}), with
solution $F_{0}^{(k)}\sim L_{k}^{-(d_{w}+2\beta)}$, $L_{k}\sim2^{k}$, as a lower bound on the
adsorption. Thus, the true $F_{0}^{(k)}$ vanishes with length $L$ as a power law with exponent
\emph{at least} as large as $d_{w}$ ($\beta=0$) but not larger than $d_{w}+2\beta\approx2.01(1)$.
Simulations of quantum walks on DSG up to $k=12$ generations shown in Fig.~\ref{fig:Extrapolation}
suggest a \emph{unique} exponent $\approx1.23(1)$ that is only minutely above $d_{w}$, independent
of the IC.

\section{How RG-Fixed Points Fail to Determine Asymptotic Properties\label{sec:RG-Fixed-Points-are}}

As Sec.~\ref{sec:RG-Flow-for-DSG} has shown, extracting the scaling
$\rho\left(x,t\right)\sim f\left(x^{d_{w}}/t\right)$ from the fixed
point proves \emph{insufficient} for the quantum walk on DSG. While
the fixed point of the RG for the quantum walk on the line naively
appears to reproduce the known scaling properties, the discussion
shows that this is likely a coincidence for this rather simple scenario.
The examples of the line in Sec.~\ref{sec:Renormalization-1dQW} and
of DSG in Sec.~\ref{sec:Renormalization-DSG} shows, the fixed point
found in both cases appears to coincide with the fractal exponent,
$d_{f}=\log_{2}\lambda_{f}$, that refers to structural properties
of the lattice, rather then to the dynamics of the walk itself. 

Since most dynamic observables require an extended examination in
the complex-$z$ plane, the example of the Sierpinski gasket suggest
that the same holds true for the RG-flow. As the data collapse in
Fig.~\ref{fig:collapseDSG} demonstrates, the way to extract the eigenvalue
$\lambda_{w}$ (and, hence, $d_{w}$) that is consistent with the
RG on fractals (such as DSG below) results from the scaling 
\begin{equation}
\tilde{\psi}_{a}^{(k)}\left(\theta\right)\sim2^{-\beta}\tilde{\psi}_{a}^{(k-1)}\left(\lambda_{w}\theta\right)\label{eq:functionalrescale}
\end{equation}
at large $k$ near a fixed point for $\theta\to0$, such that $\lambda_{w}^{k}\theta\gtrsim1$.
Rescaling $\theta$ corresponds to $z\to z^{1/\lambda_{w}}$ for $k\to k+1$,
and hence, from Eq.~(\ref{eq:LaplaceT}) we see that it amounts to
a rescaling of time $t$ with $\lambda_{w}$ when $L$ doubles, such
that $d_{w}=\log_{2}\lambda_{w}$. The scenario posed by Eq.~(\ref{eq:functionalrescale})
is depicted in Fig.~\ref{fig:RGflowPhase}, which suggests that the
RG-flow will have to be solved asymptotically in an intermediate scaling
regime, as it can never reach the fixed point for $\lambda_{w}<\lambda_{f}$.
While this has not be achieved yet, the simplicity of the exponent
suggests that such an exact analysis should be possible. It suggests
that the RG for quantum walks requires an entirely new approach, beyond
the usual fixed point analysis.

The fixed-point analysis happened to be successful for the \emph{1d}-line
because there fractal and walk exponents coincide, $d_{w}=d_{f}=d=1$
(and $\beta=0$ here), as demonstrated by the data-collapse in Fig.~\ref{fig:Fpm1dQW}(b).
As the limit $\theta\to0$ for the simple line is not singular, it
is not surprising that scaling collapse and traditional RG analysis
near the fixed point provide identical results.

\section{Conclusions\label{sec:Conclusions}}

In conclusion, we have devised a method to determine the asymptotic
behavior of discrete-time quantum walks on the dual Sierpinski gasket
using RG. Fractal graphs, as well as random networks, lack the translational
symmetry that is essential to study quantum walks on lattices. The
present treatment can be applied to renormalizable structures,\cite{Redner01}
ultimately to generate analytical results for important physical quantities
such as the spreading rate of the probability distributions. However,
compared to random walks, quantum walks require extending RG into
the complex plane, which we have explored in some detail. We confirmed
that quantum walks are more intricate than random walks, and we analyzed
the effects of geometry on quantum interference. Direct numerical
simulations support our conclusions.

The RG analysis for quantum walks appears to be more complicated than
for classical random walks, a likely result of unitarity, which precludes
the typical contractive mapping that makes RG of classical, stochastic
processes easy. Yet, our results suggest that such quantum systems
will ultimately find just as exact a description as those for classical
systems. In turn, much more will be gained, as these quantum processes
exhibit a far richer phenomenology compared to the rather structureless
diffusion process.

Finally, we note that the potential scope of RG is much broader\cite{Goldenfeld}
than merely as a toll to calculate exponents for some specific fractals,
where it happens to be exact. Once the present technical issues have
been resolved, it should be possible to use the RG, exactly or approximately,
to classify the asymptotic properties of quantum walks, and hopefully
other quantum algorithms, into universality classes. Such a classification
ultimately should serve as the basis for understanding, and hence,
control of the observed behaviors.

\section*{Acknowledgements\label{sec:acknowledgements}}

We acknowledge financial support from CNPq, LNCC, and
the U.~S.~National Science Foundation through grant DMR-1207431. SB
thanks LNCC for its hospitality and acknowledges financial support
through a research fellowship by the ``Ciencia sem Fronteiras''
program in Brazil.

\selectlanguage{english}%
\bibliography{/Users/stb/Boettcher}

\section*{Appendix\label{part:Supplement}}

\subsection{Absorption on the Line\label{sub:Absorption-for-Small}}

We consider the case of two absorbing walls on both ends of the simple
line with the initial conditions (IC) located on a single site right
next to the left wall, as shown in Fig.~\ref{fig:1dQWRG}a. It is
convenient to identify the IC-site as the origin ($i=0$), i.e., the
left-absorbing site is $\tilde{\psi}_{-}$, and the right wall is
located on site $i=2^{k}$ with $\tilde{\psi}_{+}$. Since nothing
escapes out of the absorbing sites, we have the master equations:
\begin{eqnarray}
\tilde{\psi}_{-} & = & B_{0}\tilde{\psi}_{0},\nonumber \\
\tilde{\psi}_{0} & = & N_{0}\tilde{\psi}_{0}+B_{0}\tilde{\psi}_{1}+\psi_{IC},\\
\tilde{\psi}_{x} & = & M_{0}\tilde{\psi}_{x}+A_{0}\tilde{\psi}_{x-1}+B_{0}\tilde{\psi}_{x+1},\quad\left(1\leq i\leq2^{k}-2\right),\nonumber \\
\tilde{\psi}_{2^{k}-1} & = & M_{0}\tilde{\psi}_{2^{k}-1}+A_{0}\tilde{\psi}_{2^{k}-2},\nonumber \\
\tilde{\psi}_{+} & = & A_{0}\tilde{\psi}_{2^{k}-1},\nonumber 
\end{eqnarray}
This setup has been chosen exactly such that \emph{all} quantities
renormalize according to the flow in (\ref{eq:recur1dPRWmass}), avoiding
some of the special considerations typically required near boundaries.
The only exception refers to the self-term at $\tilde{\psi}_{0}$:
Although initially $N_{0}=M_{0}=0$, its recursion is $N_{k+1}=N_{k}+B_{k}\left(I-M_{k}\right)^{-1}A_{k}$
instead. (This geometry resembles exactly the setup of Ref.~\onlinecite{Ambainis01},
with which we can now compare.)

In each recursion-step, every second intervening site still present
is eliminated, see the sequence in Fig.~\ref{fig:1dQWRG}a. As the
last setting suggests, we are left with three relations, 
\begin{eqnarray}
\tilde{\psi}_{-} & = & B_{0}\tilde{\psi}_{0},\nonumber \\
\tilde{\psi}_{0} & = & N_{k}\tilde{\psi}_{0}+\psi_{IC},\label{eq:1dQWfinal}\\
\tilde{\psi}_{+} & = & A_{k}\tilde{\psi}_{0}.\nonumber 
\end{eqnarray}
We now simply eliminate $\tilde{\psi}_{0}$ to get 
\begin{eqnarray}
\tilde{\psi}_{-}^{(k)} & = & B_{0}\left(1-N_{k}\right)^{-1}\psi_{IC},\nonumber \\
\tilde{\psi}_{+}^{(k)} & = & A_{k}\left(1-N_{k}\right)^{-1}\psi_{IC}.\label{eq:Fpm}
\end{eqnarray}

\subsection{Absorption on the Dual Sierpinski Gasket\label{sec:Quantum-Walks-DSG}}

Using the recursions developed in Eqs.~(\ref{eq:RGflowDSGPRW}),
we exactly evolve from the raw hopping coefficients $A_{0},B_{0}$,
and $M_{0}$ ($C$ does not renormalize, i.e., $C_{k}=C_{0}$) to
the $k$-th stage; the \emph{2nd}-to-final stage is shown in Fig.~\ref{fig:1dQWRG}b.
After tracing out the remaining six inner sites, we have 
\begin{eqnarray}
\tilde{\psi}_{0} & = & C_{0}\tilde{\psi}_{1},\nonumber \\
\tilde{\psi}_{1} & = & M_{k}\tilde{\psi}_{1}+A_{k}\tilde{\psi}_{3}+B_{k}\tilde{\psi}_{2},\label{eq:DSG_endgame}\\
\tilde{\psi}_{2} & = & M_{k}\tilde{\psi}_{2}+A_{k}\tilde{\psi}_{1}+\left(B_{k}+C_{0}\right)\tilde{\psi}_{3}+\frac{1}{\sqrt{2}}\psi_{IC},\nonumber \\
\tilde{\psi}_{3} & = & M_{k}\tilde{\psi}_{3}+\left(A_{k}+C_{0}\right)\tilde{\psi}_{2}+B_{k}\tilde{\psi}_{1}+\frac{1}{\sqrt{2}}\psi_{IC}.\nonumber 
\end{eqnarray}
Here, we assume symmetric IC applied at the two corner sites opposite
the absorbing wall; the procedure is easily extended to two unequal
IC. For $\eta\in\left\{ A_{k},B_{k,}C_{0},\psi_{IC}\right\} $ we
define $\bar{\eta}_{k}=\left(I-M_{k}\right)^{-1}\eta_{k}$ and make
the Ansatz 
\begin{equation}
\tilde{\psi}_{2}=X\tilde{\psi}_{1}+U\bar{\psi}_{IC},\quad\tilde{\psi}_{3}=Y\tilde{\psi}_{1}+V\bar{\psi}_{IC}.
\end{equation}
Inserting in Eqs.~(\ref{eq:DSG_endgame}) for $\tilde{\psi}_{2,3}$
determines self-consistently 
\begin{eqnarray}
X & = & \left[I-\left(\bar{A}_{k}+\bar{C}_{0}\right)\left(\bar{B}_{k}+\bar{C}_{0}\right)\right]^{-1}\left[\bar{B}_{k}+\left(\bar{A}_{k}+\bar{C}_{0}\right)\bar{A}_{k}\right],\nonumber\\
Y & = & \left[I-\left(\bar{B}_{k}+\bar{C}_{0}\right)\left(\bar{A}_{k}+\bar{C}_{0}\right)\right]^{-1}\left[\bar{A}_{k}+\left(\bar{B}_{k}+\bar{C}_{0}\right)\bar{B}_{k}\right],\nonumber\\
U & = & \left[I-\left(\bar{A}_{k}+\bar{C}_{0}\right)\left(\bar{B}_{k}+\bar{C}_{0}\right)\right]^{-1}\left[I+\left(\bar{A}_{k}+\bar{C}_{0}\right)\right],\\
V & = & \left[I-\left(\bar{B}_{k}+\bar{C}_{0}\right)\left(\bar{A}_{k}+\bar{C}_{0}\right)\right]^{-1}\left[I+\left(\bar{B}_{k}+\bar{C}_{0}\right)\right].\nonumber
\end{eqnarray}
Inserting for $\tilde{\psi}_{1}$ and then into $\tilde{\psi}_{0}$
in Eqs.~(\ref{eq:DSG_endgame}) yields 
\begin{equation}
\tilde{\psi}_{0}=C_{0}\left(I-\bar{A}_{k}X-\bar{B}_{k}Y\right)^{-1}\left(\bar{A}_{k}U+\bar{B}_{k}V\right){\bar{\psi}}_{IC}.\label{eq:WallPsiDSG}
\end{equation}

\end{document}